\documentclass[twocolumn,aps,prb,superscriptaddress,a4paper,showpacs,showkeys,lengthcheck]{revtex4-1}

\usepackage{graphicx}% Include figure files
\usepackage{dcolumn}% Align table columns on decimal point
\usepackage{bm}% bold math
\usepackage{type1cm}% scalable Adobe fonts only
\usepackage{ulem}

\usepackage{amssymb,amsmath}

\usepackage[dvips]{color}

\newcommand{\mm}[1]{\mbox{$#1$}}
\newcommand{\dstd}{\mathrm{d}}

\newcommand{\mbf}[1]{\bm{#1}}

\newcommand{\ms}{\mbox{$\mu_{\mathrm{spin}}$}}
\newcommand{\mo}{\mbox{$\mu_{\mathrm{orb}}$}}

\newcommand{\mB}{\mbox{$\mu_{B}$}}
\newcommand{\nh}{\mbox{$n_{h}$}}

\newcommand{\zcp}{\mbox{$z_{\text{Co-Pd}}$}}

\newcommand{\Led}{\mbox{$L_{2,3}$\ edge}}
\newcommand{\Ta}{\mbox{$T_{\alpha}$}}
\newcommand{\tz}{\mbox{$T_{z}$}}
\newcommand{\MM}{\mbox{$\bm{M}$}}

\newcommand{\ea}{{\it et al.}}
\newcommand{\ai}{{\it ab initio }}

\begin{document}

\title{Co monolayers and adatoms on Pd(100),
  Pd(111) and Pd(110): Anisotropy of magnetic properties}

% Please use the appropriate macro for each type of information


\author{O. \surname{\v{S}ipr}} 
\email{sipr@fzu.cz}
\homepage{http://www.fzu.cz/~sipr} \affiliation{Institute of Physics
  of the ASCR v.~v.~i., Cukrovarnick\'{a}~10, CZ-162~53~Prague, Czech
  Republic }

\author{S. \surname{Bornemann}} \affiliation{Universit\"{a}t
  M\"{u}nchen, Department Chemie, Butenandtstr.~5-13,
  D-81377~M\"{u}nchen, Germany}

\author{H. \surname{Ebert}} \affiliation{Universit\"{a}t M\"{u}nchen,
  Department Chemie, Butenandtstr.~5-13, D-81377~M\"{u}nchen, Germany}

\author{S. \surname{Mankovsky}} \affiliation{Universit\"{a}t
  M\"{u}nchen, Department Chemie, Butenandtstr.~5-13,
  D-81377~M\"{u}nchen, Germany}

\author{J. \surname{Vack\'{a}\v{r}}} \affiliation{Institute of Physics
  of the ASCR v.~v.~i., Cukrovarnick\'{a}~10, CZ-162~53~Prague, Czech
  Republic }

\author{J. \surname{Min\'{a}r}} \affiliation{Universit\"{a}t
  M\"{u}nchen, Department Chemie, Butenandtstr.~5-13,
  D-81377~M\"{u}nchen, Germany}

\date{\today}

\begin{abstract}
We investigate to what extent the magnetic properties of deposited
nanostructures can be influenced by selecting as a support different
surfaces of the same substrate material.  Fully relativistic \ai
calculations were performed for Co monolayers and adatoms on Pd(100),
Pd(111), and Pd(110) surfaces.  Changing the crystallographic
orientation of the surface has a moderate effect on the spin magnetic
moment and on the number of holes in the $d$ band, a larger effect on
the orbital magnetic moment but sometimes a dramatic effect on the
magnetocrystalline anisotropy energy (MAE) and on the magnetic dipole
term \Ta.  The dependence of \Ta\ on the magnetization
direction~$\alpha$ can lead to a strong apparent anisotropy of the
spin magnetic 
moment as deduced from the X-ray magnetic circular dichroism (XMCD)
sum rules.  For systems in which the spin-orbit coupling is not very
strong, the \Ta\ term can be understood as arising from the
differences between components of the spin magnetic moment associated
with different magnetic quantum numbers $m$.
\end{abstract}

\pacs{75.70.Ak,75.30.Gw,78.70.Dm,73.22.Dj}

\keywords{magnetism,anisotropy,nanosystems,XMCD}

\maketitle

%%%%%%%%%%%%%%%%%%%%%%%%%%%%%%%%%%%%%%%%%%%%%%%%%%%%%%%%%%%%%%%

\section{Introduction}   \label{sec-intro}

The magnetic properties of surface deposited nanostructures have been
in the ongoing focus of many experimental and theoretical
investigations as they often exhibit interesting and sometimes
unexpected phenomena.  One of the main features in this context is
that the local magnetic moments and their mutual interaction as well
as the magnetocrystalline anisotropy energy (MAE) are in general
different and often much larger in nanostructures than in
corresponding bulk systems.  Various aspects of the magnetism of many
different nanostructures were studied in the past to identify the key
factors which could then be used to tune the properties of such
systems in a desired way.  It has been known for some time that one
such key factor is the coordination number, with smaller coordination
numbers generally implying larger magnetic
moments.\cite{SKE+04,MLZ+06,BSM+12} However, coordination numbers
alone do not fully determine magnetism of nanostructures.  The
chemical composition can play a significant role as well.  An Fe
monolayer, for instance, has a larger spin magnetic moment when
deposited on Au(111) than when deposited on Pt(111), whereas for a Co
monolayer it is {\it vice versa}.\cite{BSM+12} The situation is even
more diverse for the MAE where different substrates may lead to
different properties of systems of otherwise identical geometries.
For example, Co$_{2}$ and Ni$_{2}$ dimers on Pt(111) have out-of-plane
magnetic easy axis but the same dimers on Au(111) have an in-plane
magnetic easy axis.\cite{BSM+12}

Experimental research on magnetism of nanostructures relies heavily on
the X-ray magnetic circular dichroism (XMCD) sum
rules.\cite{GDM+02,LFN+02,GRV+03} The strength of these sum rules is
that they give access to spin magnetic moments \ms\ and orbital
magnetic moments \mo\ separately and in a chemically specific
way.\cite{CTAW93,TCSvdL92} However, the XMCD spin sum rule does not
provide \ms\ alone but only its combination
\mm{\mu_{\text{spin}}+7T_{\alpha}}, where \Ta\ is the magnetic dipole
term (for the magnetization $\mbf{M}$ parallel to the $\alpha$
axis, $\alpha$=$x,y,z$).\cite{CTAW93} For bulk systems, \Ta\ can be
usually neglected 
but for surfaces and clusters the \Ta\ term can have significant
influence, as it has been demonstrated
experimentally\cite{GDG+02,SMC+10} and
theoretically.\cite{WF94,KEDF02,SME09b} The anisotropy of the magnetic
dipole term was predicted on general grounds\cite{SK95a} and some
estimates concerning the magnitude of this anisotropy in non-cubic
bulk systems were given based on atomic-like model
Hamiltonians\cite{SK95a} or on \ai calculations.\cite{KEF04}

Magnetic nanostructures may be prepared by combining and arranging
different magnetic elements on different substrates.  In this respect
one can also address surfaces of different crystallographic
orientations.  Thus, it is important to know how the magnetic
properties can be controlled by selecting for the substrate
crystallographically different surfaces of the same material and
whether one can expect different effects for complete monolayers and
for adatoms.  Connected with this is the question about the effects on
the \Ta\ term, because XMCD is perhaps the most frequently used
experimental technique in this field and it is desirable to know how
\Ta\ can influence the values of magnetic moments deduced from the
XMCD sum rules. For planning and interpreting such experiments, it
would be very useful not only to know the \Ta\ values from \ai
calculations but also to have a simple intuitive interpretation of the
\Ta\ term.

In order to learn more about this, we undertook a systematic study of
Co monolayers and adatoms on Pd(100), Pd(111), and Pd(110) surfaces.
Fully relativistic \ai calculations were performed to obtain \ms, \mo,
and \Ta\ for different magnetization directions.  The MAE was
determined for all these systems as well.  The accuracy of an
approximative expression for the \Ta\ term was checked to see whether
it captures the essential physics.  It is shown in the following that
monolayers and adatoms on different crystallographic surfaces may have
indeed quite different magnetic properties, especially as concerns the
MAE.  Moreover, it is also demonstrated how the dependence of the
\Ta\ term on the magnetization direction leads to a surprisingly
strong apparent anisotropy of \ms\ as deduced from the XMCD sum rules.

%%%%%%%%%%%%%%%%%%%%%%%%%%%%%%%%%%%%%%%%%%%%%%%%%%%%%%%%%%%%%%%

\section{Methods}

\subsection{Investigated systems}   \label{sec-systems}
 
% Figure planned for 1 1/2 column
\begin{figure*}
\includegraphics[viewport=0.1cm 0.1cm 12.8cm 4.2cm]{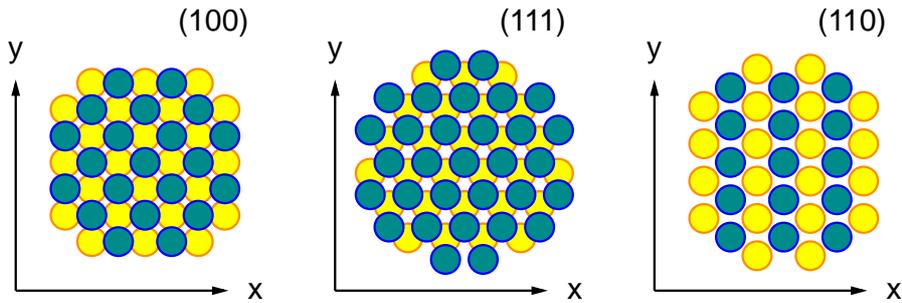}%
\caption{(Color online) Structure diagrams for a Co monolayer on
  Pd(100), Pd(111) and Pd(110). The blue and yellow circles represent
  the Co and Pd atoms, respectively. The orientation of the $x$ and
  $y$ coordinates used throughout this paper is also shown.}
\label{fig-surfplot}
\end{figure*}

We investigated Co monolayers on Pd(100), Pd(111) and Pd(110) and also
Co adatoms on the same surfaces. The corresponding structure diagrams
are shown in Fig.\ \ref{fig-surfplot} (for adatoms, obviously only one
Co atom is kept).  Two hollow adatom positions are possible for the
(111) surface, differing by the position of the adatom with respect to
the sub-surface layer; we consider the fcc position in this work
(unless specified otherwise).

The Pd substrate has fcc structure with lattice constant $a$=3.89~\AA.
To determine the distances between the Co atoms and the substrate, we
relied in most cases on the ``constant volume approximation'': the
vertical Co--Pd interplanar distance $z_{\text{Co-Pd}}$ is taken as an
average between the interlayer distance in bulk Pd and the interlayer
distance in a hypothetical pseudomorphically grown fcc Co film
compressed vertically in such a way that the atomic volume of Co is
the same as in bulk Co.\cite{KTM+00} In addition we took also into
account relevant experimental data and results of ab-initio geometry
relaxations when available. For example, the constant volume
approximation yields $z_{\text{Co-Pd}}$=1.70~\AA\ for a Co monolayer
on Pd(100) while we took $z_{\text{Co-Pd}}$=1.65~\AA\ instead,
following the surface X-ray diffraction experiment of Meyerheim
\ea\cite{MPE+07} For the other two surfaces we used the constant
volume approximation distances, namely,
$z_{\text{Co-Pd}}$=1.96~\AA\ for Co on Pd (111) and
$z_{\text{Co-Pd}}$=1.20~\AA\ for Co on Pd(110).  In the case of the
(111) surface we can compare our distance with an EXAFS-derived
experimental distance
$z_{\text{Co-Pd}}$=2.02~\AA\ (Ref.\ \onlinecite{MMN+07}) and with an
\ai equilibrium distance
$z_{\text{Co-Pd}}$=1.91~\AA\ (Ref.\ \onlinecite{WLF+99}).  It follows
from this comparison that the constant-volume-approximation leads to
reasonable distances.

Systems with interplanar distances as given above will be called
systems with ``optimized geometries''.  Apart from that, we
investigate for comparison also systems where the Co atoms are located
in ideal positions of the underlying Pd lattice.  For this we use the
designation ``bulk-like geometry''.  The interplanar distances are
summarized in Tab.\ \ref{tab-geom}.

\begin{table}
\caption{Vertical distances $z_{\text{Co-Pd}}$ between the plane
  containing Co atoms and plane containing Pd atoms for systems
  investigated in this study.  The unit is \AA.}
\label{tab-geom}
\begin{ruledtabular}
\begin{tabular}{ldd}
  \multicolumn{1}{c}{surface}   &  
  \multicolumn{1}{c}{optimized geometry} & 
  \multicolumn{1}{c}{bulk-like geometry} \\
(100)  &  1.65  &  1.95 \\
(111)  &  1.96  &  2.25 \\
(110)  &  1.20  &  1.38 \\
\end{tabular}
\end{ruledtabular}
\end{table}

For adatoms we use the same $z_{\text{Co-Pd}}$ distances as for
monolayers.  This is a simplification because the constant volume
approximation will work worse for adatoms than for monolayers. For
example the \ai $z_{\text{Co-Pd}}$ distance for a Co adatom on Pd(111)
is 1.66~\AA\ (Ref.\ \onlinecite{BLD+10}) in contrast to our optimized
geometry value of 1.96~\AA.  However, by using identical
\zcp\ distances for monolayers and adatoms, the net effect due to the
change in Co coordination can be studied.  It will be shown that the
effect of varying the distances is in fact smaller than the effect of
monolayer-to-adatom transition.

\subsection{Computational scheme}   \label{sec-comput}

The calculations were performed within the \ai spin density functional
framework, relying on the local spin density approximation (LSDA) with
the Vosko, Wilk and Nusair parametrization for the exchange and
correlation potential.\cite{VWN80} The electronic structure is
described, including all relativistic effects, by the Dirac equation,
which is solved using the spin polarized relativistic
multiple-scattering or Korringa-Kohn-Rostoker (SPR-KKR) Green's
function formalism\cite{EKM11} as implemented in the {\sc spr-tb-kkr}
code.\cite{tbkkr-code} The potentials were treated within the atomic
sphere approximation (ASA) and for the multipole expansion of the
Green's function, an angular momentum cutoff
\mm{\ell_{\mathrm{max}}}=3 was used.

The electronic structure of Co monolayers on Pd surfaces was
calculated by means of the tight-binding or screened KKR
technique.\cite{ZDU+95} The substrate was modeled by slabs of 13--14
layers (i.e.\ a thickness of 17--27~\AA, depending on the surface
orientation), the vacuum was represented by 4--5 layers of empty
sites.  The adatoms were treated as embedded impurities: first the
electronic structure of the host system (clean surface) was calculated
and then a Dyson equation for an embedded impurity cluster was
solved.\cite{BMP+05} The impurity cluster contains 135 sites if not
specified otherwise; this includes a Co atom, 50--60 Pd atoms and the
rest are empty sites.

It should be stressed that the embedded clusters define the region
where the electronic structure and potential of the host is allowed to
relax due to the presence of the adatom and {\it not} the size of the
considered system.  In this respect the Green's function approach
differs from the often used supercell approach: there is an
unperturbed host beyond the relaxation zone in the former approach
while in the latter approach, the supercell is terminated either by
vacuum or by another (interfering) relaxation zone pertaining to an
adjacent adatom.  The sizes of the embedded clusters and the sizes of
the supercells thus have a different meaning and cannot be directly
compared.

The magnetocrystalline anisotropy energy (MAE) is calculated by means
of the torque $T^{(\hat{n})}_{\hat{u}}$ which describes the variation
of the energy if the magnetization direction $\hat{n}$\ is
infinitesimally rotated around an axis $\hat{u}$. For uniaxial systems
where the total energy can be approximated by
\[
E(\theta) = E_{0} \, + \, K_{2} \sin^{2}(\theta) \, + \,
K_{4} \sin^{4}(\theta) \; ,
\]
the difference \mm{E(90^{\circ})-E(0^{\circ})}\ is equal to the torque
evaluated for \mm{\theta=45^{\circ}}.\cite{WWW+96} The torque itself
was calculated by relying on the magnetic force theorem.\cite{SSB+06}

Apart from the magnetocrystalline anisotropy induced by the spin-orbit
coupling, the magnetic easy axis is also determined by the so-called
shape anisotropy caused by magnetic dipole-dipole interactions.  The
shape anisotropy energy is usually evaluated classically by a lattice
summation over the magnetostatic energy contributions of individual
magnetic moments, even though it can be in principle obtained \ai via
a Breit Hamiltonian.\cite{BMB+12} In this paper, we always deal only
with the magnetocrystalline contribution to the magnetic anisotropy
unless stated otherwise.

%%%%%%%%%%%%%%%%%%%%%%%%%%%%%%%%%%%%%%%%%%%%%%%%%%%%%%%%%%%%%%%

\section{Results}   \label{sec-res}

%--%--%--%--%--%--%--%--%--%--%--%--%--%--%--%--%--%--%--%--%--

\subsection{Magnetic moments and magnetocrystalline anisotropy} 

\label{sec-moms}

% Wide table planned for two columns
\begin{table*}
\caption{Magnetic properties of Co monolayers and adatoms on Pd(100),
  Pd(111), and Pd(110).  The first column specifies whether the values
  are for a monolayer or for an adatom, the second column contains
  spin magnetic moment for the Co atom for \mm{\mbf{M}\| z} (in units
  of \mB), the third column contains number of holes in the $d$ band
  for \mm{\mbf{M}\| z}.  The fourth, fifth and sixth columns contain
  orbital magnetic moments for the Co atom for \mm{\mbf{M}\| z},
  \mm{\mbf{M}\| x}, and \mm{\mbf{M}\| y}, respectivelly.  The last
  three columns contain the MAE between indicated magnetization
  directions (in meV per Co atom). Numbers without brackets stand for
  systems with optimized Co--Pd distances, numbers in brackets stand
  for systems with a bulk-like geometry (see
  Sec.\ \ref{sec-systems}).}
\label{tab-moms}
\begin{ruledtabular}
\begin{tabular}{ldddddddd}
   & 
    \multicolumn{1}{c}{$\mu_{\text{spin}}^{(z)}$} & 
    \multicolumn{1}{c}{$n_{h}^{(z)}$} & 
    \multicolumn{1}{c}{$\mu_{\text{orb}}^{(z)}$} & 
    \multicolumn{1}{c}{$\mu_{\text{orb}}^{(x)}$} & 
    \multicolumn{1}{c}{$\mu_{\text{orb}}^{(y)}$} & 
    \multicolumn{1}{c}{$E^{(x)}-E^{(z)}$} & 
    \multicolumn{1}{c}{$E^{(y)}-E^{(z)}$} & 
    \multicolumn{1}{c}{$E^{(x)}-E^{(y)}$} \\
%%%\hline  \\  [-3.5ex]
\hline 
  \multicolumn{8}{c}{Co on Pd(100)} \\ 
monolayer & 
 2.09 &      2.45 &
     0.132 & 0.203  &   & 
       -0.73  &  & \\
          &
 (2.07) &    (2.39) &
     (0.190) & (0.241)  &   & 
       (-0.69) &  &  \\  [0.5ex]
adatom &
 2.29 &    2.57  &
     0.299 & 0.279 & &
          0.26 &  &  \\
          &
 (2.32) &   (2.53)  &
     (0.610) & (0.473) &  & 
         (2.69)  &  &  \\  [1.5ex]
  \multicolumn{8}{c}{Co on Pd(111)} \\ 
monolayer & 
 2.02  &   2.43  &
     0.135  & 0.136  &  & 
        0.36  &   & \\
     & 
 (1.99)  &   (2.41)  &
     (0.154) & (0.176) &  &
       (0.21)  &    &  \\  [0.5ex]
adatom & 
 2.35  &   2.62  &
     0.605  & 0.355  &  &
        5.50  &  &  \\
       & 
 (2.34)  &  (2.52) &
     (0.780) & (0.575)  &  & 
       (6.38)  &  &  \\ [1.5ex]
  \multicolumn{8}{c}{Co on Pd(110)} \\ 
monolayer &
 2.15 &   2.50  &
     0.192  & 0.183  &  0.210  & 
        -0.15  &  -0.43  & 0.28 \\
          & 
 (2.18)  &    (2.54)  &
     (0.215)  & (0.220)  &  (0.289)  & 
          (-0.48)  &  (-0.97) & (0.49)  \\  [0.5ex]
adatom & 
 2.20  &    2.49  &
     0.270  & 0.347  &  0.201  &
          -1.51  &  1.10  & -2.61 \\
        & 
 (2.25)  &    (2.47)  & 
     (0.349)  & (0.472)  &  (0.255)  &
          (-1.88)  &  (2.01)  & (-3.89) \\
\end{tabular}
\end{ruledtabular}
\end{table*}

To assess the effect of selecting different crystallographic surfaces
and of going from a monolayer to an adatom, we calculated magnetic
moments, numbers of holes in the Co $d$ band and the MAE for all these
systems.  The results are summarized in Tab.\ \ref{tab-moms}.  For
each system, the data are shown first for the optimized geometry and
then for the bulk-like geometry (numbers in the brackets).  The $x$,
$y$, and $z$ superscripts in the column header labels indicate the
direction of the magnetization \MM.

The spin magnetic moment \ms\ and the number of holes in the $d$ band
\nh\ are shown only for \mm{\mbf{M} \| z}, because they are
practically independent on the magnetization direction: by varying it,
\ms\ can be changed by no more than 0.2~\% and \nh\ by no more than
0.1~\%.  On the other hand, for \mo\ the differences can be quite
large.  The second in-plane magnetization direction \mm{\mbf{M} \| y}
was investigated only for the (110) surface, because there is only
very small ``intraplanar anisotropy'' for the (100) and (111) surfaces
(this issue is addressed in more detail in Sec.\ \ref{sec-azim}).  For
bulk hcp Co we get \ms=1.61~\mB, \mo=0.08~\mB\ and \nh=2.48.

Changing the surface orientation has a moderate effect on \ms\ and
\nh.  The differences in \ms\ when going from one surface to another
are at most 9~\%.  For \nh\ these differences are at most 5~\%.
However, the situation is quite different for \mo\ where the
differences are 20--50~\%.  The sensitivity in \mo\ finds its
counterpart in the sensitivity of the MAE.  For example, the magnetic easy axis
for a Co monolayer is in-plane for the (100) and
(110) surfaces but out-of-plane for the (111) surface.  For the
adatom, the easy axis is in-plane for the (110) surface but
out-of-plane for the (100) and (111) surfaces.  So in this respect the
choice of the crystallographic surface can have a dramatic influence.

Another finding emerging from Tab.\ \ref{tab-moms} is that as concerns
\ms, the difference between monolayers and adatoms is only
quantitative in most cases.  A surprisingly small difference in this
respect is found for the (110) surface.  As the same Co--Pd distances
have been used for monolayers and adatoms, one observes here the net
effect of the change in Co coordination.  For \mo, the difference
between monolayers and adatoms is obviously much larger than for \ms.
For the MAE this difference can again be essential: The magnetic easy
axis for a Co monolayer on Pd(100) is in-plane while for a Co adatom
on the same surface it is out-of-plane.  Similarly, the magnetic easy axis 
for a monolayer on Pd(110) is parallel to the
$y$-axis while for an adatom it is parallel to the $x$-axis.

Changing the distance between Co atoms and the surface clearly affects
the magnetic properties (cf.\ the values with and without brackets in
Tab.\ \ref{tab-moms}).  However, it is noteworthy that the effect of
geometry relaxation is smaller than the effect of the transition from
the monolayer to the adatom.

We calculated also the magnetic shape anisotropy for the monolayers
(classically, via a lattice summation, taking into account also
moments on Pd atoms).  As expected, this contribution favors always an
in-plane orientation of the magnetization.  For Co monolayers on
Pd(100) and Pd(111), we get
$E_{\text{dip-dip}}^{(x)}-E_{\text{dip-dip}}^{(z)}=-0.1$~meV.  For Co
monolayers on Pd(110), there is a small difference regarding the $x$
and $y$ directions: we get
$E_{\text{dip-dip}}^{(x)}-E_{\text{dip-dip}}^{(z)}=-0.07$~meV and
$E_{\text{dip-dip}}^{(y)}-E_{\text{dip-dip}}^{(z)}=-0.09$~meV.  By
comparing these values with the values shown in Tab.\ \ref{tab-moms},
we see that the shape anisotropy energy is smaller in magnitude than
the magnetocrystalline anisotropy energy and thus the shape anisotropy
does not change the orientation of the magnetic easy axis as
determined by the magnetocrystalline anisotropy.

%--%--%--%--%--%--%--%--%--%--%--%--%--%--%--%--%--%--%--%--%--

\subsection{Induced magnetic moments}     \label{sec-induced}

\begin{table}
\caption{Spin magnetic moments for Pd atoms which are first, second
  and third nearest neighbors of Co atoms, in units of \mB.  As in
  Tab.\ \protect\ref{tab-moms}, the numbers without brackets stand for
  systems with optimized geometry and the numbers in brackets stand
  for systems with bulk-like geometry.}
\label{tab-induced}
\begin{ruledtabular}
\begin{tabular}{lddd}
   & 
    \multicolumn{1}{c}{Pd(1)} & 
    \multicolumn{1}{c}{Pd(2)} & 
    \multicolumn{1}{c}{Pd(3)}  \\
%%%\hline  \\  [-3.5ex]
\hline 
  \multicolumn{4}{c}{Co on Pd(100)}    \\ 
monolayer &  0.29    &  0.17   &  0.11     \\
          &  (0.25)  &   (0.16)  & (0.10)  \\  [0.5ex]
adatom    &  0.18    &   0.06    & 0.04  \\
          &  (0.15)  &   (0.06)  &  (0.04) \\  [1.5ex]
  \multicolumn{4}{c}{Co on Pd(111)} \\ 
monolayer &  0.32    &  0.16     &  0.03    \\
          &  (0.25)  &   (0.15)  &  (0.06)  \\  [0.5ex]
adatom    &  0.16    &   0.02    &  0.04    \\
          &  (0.12)  &   (0.02)  &  (0.03)  \\  [1.5ex]
  \multicolumn{4}{c}{Co on Pd(110)} \\ 
monolayer &  0.29    &  0.22   &  0.17     \\
          &  (0.29)  &   (0.24)  & (0.19)  \\  [0.5ex]
adatom    &  0.15    &   0.04    & 0.04  \\
          &  (0.15)  &   (0.05)  &  (0.04) \\  
\end{tabular}
\end{ruledtabular}
\end{table}

Palladium is not magnetic as an element but it is quite
polarizable.\cite{Zel93,PMS+10} Spin magnetic moments induced in the
Pd substrate by Co monolayers and adatoms are shown in
Tab.\ \ref{tab-induced} for all three surface orientations.  In the
case of Co monolayers, the induced moments are shown for the first
three atomic layers of Pd below the Co layer [denoted as Pd(1), Pd(2)
  and Pd(3) in Tab.\ \ref{tab-induced}].  Note that the interlayer
distances are 1.95~\AA, 2.25~\AA\ and 1.38~\AA\ for the (100), (111)
and (110) surfaces, respectively.  The relatively large \ms\ for the
Pd(2) and Pd(3) sites in the case of the (110) surface reflects the
relatively small interlayer distance for this crystallographic
orientation.

In the case of adatoms, the description is formally more complicated
because Pd atoms belonging to the same coordination shell around the
Co atom are not all equivalent: some of them belong to the surface
layer, some to the sub-surface layer and so on.  In order not be
overwhelmed by too much data, we display here only moments averaged
over all atoms of a given coordination shell.  Symbols Pd(1), Pd(2),
and Pd(3) in Tab.\ \ref{tab-induced} stand now for the first, second,
and third shell of Pd atoms around the Co adatom.

Moreover, we also calculated the orbital magnetic moments for the Pd
atoms in all systems and we found that \mo\ amounts to about 8--17~\%
of the corresponding \ms.

In this section we deal only with magnetic moments on those Pd atoms
which are close to the Co atoms.  The issue of more distant Pd atoms
and of the total charge contained in the polarization cloud is dealt
with in the Appendix.  Here, we would only like to stress that it
follows from the analysis outlined in the Appendix that our model
system is clearly adequate to yield reliable values of induced
magnetic moments for the Pd(1), Pd(2), and Pd(3) sites.

%--%--%--%--%--%--%--%--%--%--%--%--%--%--%--%--%--%--%--%--%--

\subsection{Azimuthal dependence of MAE}     \label{sec-azim}

% Figure planned for 1 1/2 column

\begin{figure*}
\includegraphics[viewport=0.0cm 0.0cm 17.7cm 6.0cm]{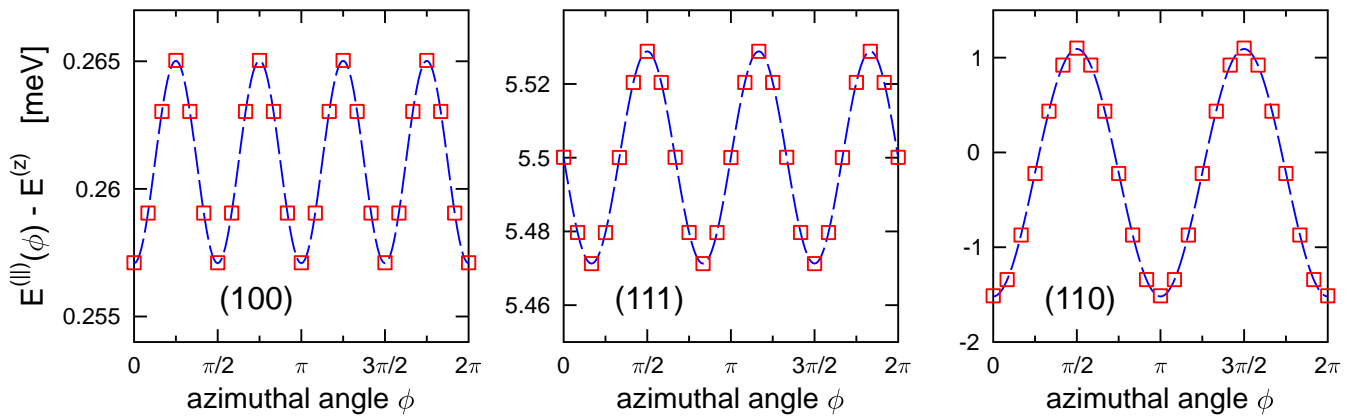}%
\caption{(Color online) Difference between total energies for in-plane
  and out-of-plane magnetization for a Co adatom on Pd (100), (111),
  and (110) surfaces (bulk-like geometry).  Points are results of the
  calculation, dashed lines are sinusoidal fits.  The orientation of
  the $x$ and $y$ axes is as in Fig.\ \protect\ref{fig-surfplot}.}
\label{fig-azimuth}
\end{figure*}

In general, the MAE defined as the difference between total energies
for in-plane and out-of-plane orientation of the magnetization will
depend on the azimuthal angle $\phi$.  This dependence is often
ignored but may sometimes be significant.  In our case, the
intraplanar MAE \mm{E^{(x)}-E^{(y)}}\ is quite comparable to
\mm{E^{(x)}-E^{(z)}}\ or \mm{E^{(y)}-E^{(z)}} for the (110) surface
(see Tab.~\ref{tab-moms}).  To get a more complete picture, we inspect
the azimuthal dependence of \mm{E^{(\|)}(\phi) - E^{(z)}}, where
\mm{E^{(\|)}(\phi)}\ is the total energy if \MM\ is in the surface
plane ($\theta$=0$^{\circ}$) with the azimuthal angle $\phi$.  Our
results for a Co adatom on all three Pd surfaces are shown in
Fig.\ \ref{fig-azimuth}.  The data reported here were obtained for the
bulk-like geometry but the trends would be similar for any
\zcp\ distance.

One can see from Fig.\ \ref{fig-azimuth} that the \mm{E^{(\|)}(\phi) -
  E^{(z)}}\ curves follow the symmetry of the appropriate surface, as
expected.  The amplitude of these curves is the most interesting
information here.  For high-symmetry surfaces, it is almost
negligible: 0.008~meV or 3~\% of the average value for Co on Pd(100)
and 0.06~meV or 1~\% of the average value for Co on Pd(111).  For the
(110) surface, however, the amplitude is 2.6~meV and to speak about an
average MAE does not make sense in this case, as illustrated by the
fact that the magnetic easy axis is in-plane for $\phi=0^{\circ}$ and
out-of-plane for $\phi=90^{\circ}$.

%--%--%--%--%--%--%--%--%--%--%--%--%--%--%--%--%--%--%--%--%--

\subsection{Relation between magnetic dipole term and $m$-decomposed 
  spin magnetic moment}     \label{sec-tzterm}

The spin magnetic moment sum rule for the \Led\ XMCD spectra can be
written for a sample magnetized along the $\alpha$ direction
as\cite{CTAW93}
\begin{equation}
 \frac{3}{I} \, \int \left( \Delta \mu_{{L}_3}
        -2\Delta \mu_{{L}_2} \right) \, \mathrm{d}E  \, = \, 
        \frac{\mu_{\text{spin}} + 7T_{\alpha}}{n_{h}}
   \enspace ,
\label{eq-spin}
\end{equation}
where \mm{\Delta \mu_{{L}_{2,3}}} are the differences
\mm{\Delta\mu=\mu^{(+)}-\mu^{(-)}} between absorption coefficients for
the left and right circularly polarized light propagating along the
$\alpha$ direction, $I$\ is the integrated isotropic absorption
spectrum, $\mu_{\mathrm{spin}}$\ is the local spin magnetic moment
(only its $d$\ component enters here), $n_{h}$\ is the number of holes
in the $d$\ band, and $T_{\alpha}$ is the magnetic dipole term related
to the $d$ electrons. \Ta\ can be written as\cite{Sto95,Sto99}
\begin{align}
T_{\alpha} & \: = \: - \frac{\mu_{B}}{\hbar} \, 
                     \langle \hat{T}_{\alpha} \rangle  
\;\; , \notag  \\
   & \: = \:  -\frac{\mu_{B}}{\hbar} \, 
   \left\langle \, \sum_{\beta} Q_{\alpha \beta} S_{\beta} \, 
      \right\rangle 
\enspace ,
 \raisetag{1.0\baselineskip}  \label{exact}
\end{align}
where 
\begin{equation}
 Q_{\alpha \beta} \: = \: \delta_{\alpha \beta} \, - \, 
3 r^{0}_{\alpha} r^{0}_{\beta} 
\end{equation}
is the quadrupole moment operator and $S_{\alpha}$ is the spin
operator.  If $z$ is the quantization axis, the eigenvalues of $S_{z}$
are $\pm (1/2) \hbar$.

A more transparent expression for $T_{\alpha}$ can be obtained if the
spin-orbit coupling can be neglected.  Then one can write\cite{Sto99}
\begin{equation}
\begin{aligned}
\hat{T}_{x} & \: = \: \left( -\frac{\mu_{B}}{\hbar} \right) 
                        \hat{Q}_{xx} \, \hat{S}_{\bar{z}} & 
\text{for} \quad \mbf{M} & \| x 
\;\; ,\\
\hat{T}_{y} & \: = \: \left( -\frac{\mu_{B}}{\hbar} \right) 
                        \hat{Q}_{yy} \, \hat{S}_{\bar{z}} & 
\text{for} \quad \mbf{M} & \| y 
\;\; ,\\
\hat{T}_{z} & \: = \: \left( -\frac{\mu_{B}}{\hbar} \right) 
                        \hat{Q}_{zz} \, \hat{S}_{\bar{z}} & 
\text{for} \quad \mbf{M} & \| z 
\;\; ,
\end{aligned}
\end{equation}
where $\hat{Q}_{xx}$, $\hat{Q}_{yy}$ and $\hat{Q}_{zz}$ are quadrupole
moment components referred to the crystal (global) reference frame and
$\hat{S}_{\bar{z}}$ is the spin component with respect to the local
reference frame in which $\bar{z}$ is identical to the spin
quantization axis.  We are interested in the expectation value of the
$\hat{T}_{\alpha}$ operator acting on the $d$ components of the wave
function in the vicinity of the photoabsorbing site. Using for the
sake of clarity a simplified two-component formulation instead of the
full Dirac approach, the wave function can be expanded in the
angular-momentum basis as
\begin{equation}
\psi_{E\mbf{k}}(\mbf{r}) \: = \:
\sum_{\ell m} \sum_{s} a_{E \mbf{k} \ell m}^{(s)} (r) \,
Y_{\ell m} (\mbf{\hat{r}}) \, \chi^{(s)} 
\end{equation}
to obtain
\begin{equation}
T_{\alpha} \: = \:  \left( -\frac{\mu_{B}}{\hbar} \right) \,
 \int^{E_{F}}_{-\infty} \!\! \dstd E \, 
 \int_{\text{BZ}} \!\! \dstd \mbf{k} \:  
 \langle \psi_{E\mbf{k}} | 
\hat{Q}_{\alpha \alpha} \hat{S}_{\bar{z}}
 |  \psi_{E\mbf{k}} \rangle 
\;\; .
\end{equation}
Restricting ourselves just to the $\ell=2$ component and omitting the
corresponding subscript in $a_{E \mbf{k} \ell m}^{(s)} (r)$, we get
\begin{widetext}
\begin{align}
T_{\alpha} \: = \: & \left( -\frac{\mu_{B}}{\hbar} \right) \,
 \int^{E_{F}}_{-\infty} \!\! \dstd E \, 
 \int_{\text{BZ}} \!\! \dstd \mbf{k} \, 
\sum_{m m'} \sum_{s s'} \,
\int \! \dstd \mbf{r} \,  
a_{E \mbf{k} m}^{(s)\, \ast}(r) \, Y_{2m}^{\ast}(\mbf{\hat{r}})
\, Q_{\alpha \alpha} \, 
a_{m' \mbf{k} E}^{(s')}(r) \, Y_{2m'}(\mbf{\hat{r}}) 
 \, 
\langle \chi^{(s)} | \hat{S}_{\bar{z}} |  \chi^{(s')} \rangle
 \notag  \\
   \: = \: & \left( -\frac{\mu_{B}}{\hbar} \right) \, 
 \int^{E_{F}}_{-\infty} \!\! \dstd E \, 
 \int_{\text{BZ}}  \!\! \dstd \mbf{k} \, 
\sum_{m m'} \,
\int \! r^{2} \dstd r  
\left[ 
a_{E \mbf{k} m}^{\uparrow\, \ast}(r) \, a_{m' \mbf{k} E}^{\uparrow}(r)  
 \, - \, 
a_{E \mbf{k} m}^{\downarrow\, \ast}(r) \, a_{m' \mbf{k} E}^{\downarrow}(r)  
\right]  
 \times \notag \\  &  
\, 
\langle Y_{2m} | \hat{Q}_{\alpha \alpha} | Y_{2m'} \rangle 
\,  \frac{1}{2} \hbar
 \notag  \\
  \: = \: & 
\frac{1}{2} \, (-\mu_{B}) \sum_{m m'} 
\left[ N^{\uparrow}_{m m'} \, - \,  N^{\downarrow}_{m m'} \right] 
\, \langle Y_{2m} |  \hat{Q}_{\alpha \alpha} | Y_{2m'} \rangle
\;\; ,
\label{taa}
\end{align}
\end{widetext}
where the spin-dependent number of states matrix $N^{(s)}_{m m'}$ is
defined as
\[
N^{(s)}_{m m'} \: = \: 
\int^{E_{F}}_{-\infty} \!\! \dstd E \, \int_{BZ} \!\! \dstd \mbf{k} \,
\int \! r^{2} \dstd r \,
a_{E \mbf{k} m}^{(s)\, \ast}(r) \, a_{m' \mbf{k} E}^{(s)}(r) 
\;\; .
\]
The difference of the diagonal terms of $N^{(s)}_{m m}$ is just the
spin magnetic moment decomposed according to the magnetic quantum
number $m$,
\[
(-\mu_{B}) \, 
\left( N^{\uparrow}_{m m} \, - \, N^{\downarrow}_{m m} \right)
\: = \:
\mu_{\text{spin}}^{(m)}
\;\; ,
\]
with the sum of all the $m$ components giving the total spin magnetic
moment (of the $d$ electrons, in our case).  Therefore, if it was
possible to restrict the sum (\ref{taa}) just to the terms diagonal in
$m$, one would have
\begin{equation}
T_{\alpha} \: = \: \frac{1}{2} \, 
\sum_{m} \mu_{\text{spin}}^{(m)} \, 
\langle Y_{2m} |  \hat{Q}_{\alpha \alpha} | Y_{2m} \rangle
\;\; .
\label{tzspin}
\end{equation}
The procedure we employed above is essentially the one suggested by
St\"{o}hr,\cite{Sto95,Sto99} but we present it here in a more explicit
way.

\begin{table}
\caption{Diagonal components of the quadrupole operator in the basis
  of real spherical harmonics.  Non-diagonal components are all zero
  except for the components given in Eq.~(\protect\ref{nondiag}). }
\label{tab-qaa}
\begin{ruledtabular}
\begin{tabular}{lrrr}
   &   \multicolumn{1}{c}{$Q_{xx}$}  &
       \multicolumn{1}{c}{$Q_{yy}$}  &
       \multicolumn{1}{c}{$Q_{zz}$}   \\
$\langle Y_{xy} |  \hat{Q}_{\alpha \alpha} | Y_{xy} \rangle$  &
   $-\frac{2}{7}$  &  $-\frac{2}{7}$  &  $\frac{4}{7}$   \\
$\langle Y_{yz} |  \hat{Q}_{\alpha \alpha} | Y_{yz} \rangle$  &
   $\frac{4}{7}$   &  $-\frac{2}{7}$  &  $-\frac{2}{7}$  \\
$\langle Y_{3z^{2}-r^{2}} |  \hat{Q}_{\alpha \alpha} | Y_{3z^{2}-r^{2}} \rangle$ &
   $\frac{2}{7}$  &  $\frac{2}{7}$  &  $-\frac{4}{7}$   \\
$\langle Y_{xz} |  \hat{Q}_{\alpha \alpha} | Y_{xz} \rangle$  &
   $-\frac{2}{7}$  &  $\frac{4}{7}$   &  $-\frac{2}{7}$  \\
$\langle Y_{x^{2}-y^{2}} |  \hat{Q}_{\alpha \alpha} | Y_{x^{2}-y^{2}} \rangle$ &
   $-\frac{2}{7}$  &  $-\frac{2}{7}$  &  $\frac{4}{7}$   \\
\end{tabular}
\end{ruledtabular}
\end{table}

The coefficients $\langle Y_{2m} | \hat{Q}_{\alpha \alpha} | Y_{2m'}
\rangle$ can be obtained by analytic integration.  If we use the basis
of {\it real} spherical harmonics, the only ``cross-terms'' which are
non-zero are
\begin{equation}
\begin{split}
\langle Y_{x^{2}-y^{2}} |  \hat{Q}_{xx} | Y_{3z^{2}-r^{2}} \rangle 
& = (2/7) \sqrt{3}
\;\; ,
 \\
\langle Y_{x^{2}-y^{2}} |  \hat{Q}_{yy} | Y_{3z^{2}-r^{2}} \rangle 
& = -(2/7) \sqrt{3}
\;\; .
\end{split}
\label{nondiag}
\end{equation}
Otherwise, only the diagonal terms 
\begin{math}
\langle Y_{2m} |  \hat{Q}_{\alpha \alpha} | Y_{2m} \rangle
\end{math}
are non-zero and we list them in Tab.~\ref{tab-qaa} (see also
Refs.~\onlinecite{SK95a,Sto95}).  Therefore, {\it in the absence of
  spin-orbit coupling,} Eq.~(\ref{tzspin}) presents an exact
expression for $T_{z}$ and an approximate expression for $T_{x}$ and
$T_{y}$ [due to the existence of non-diagonal terms (\ref{nondiag})].
As argued by St\"{o}hr,\cite{Sto99} the non-diagonal terms drop out of
the sum in Eq.~(\ref{taa}) for high symmetry systems.

Eq.~(\ref{tzspin}) together with Tab.~\ref{tab-qaa} illustrate the
common statement that the magnetic dipole term $T_{\alpha}$ is related
to spin anisotropy: if the $m$-components of \ms\ are all identical,
$T_{\alpha}$ is zero (in the absence of spin-orbit coupling).  It is
also evident from Eq.\ (\ref{tzspin}) and Tab.\ \ref{tab-qaa} that the
\Ta\ term will generally depend on the magnetization direction
$\alpha$.

% Wide table planned for two columns
\begin{table*}
\caption{Spin magnetic moment decomposed according to the magnetic
  quantum number $m$ together with the corresponding
\begin{math}
T^{(m)}_{\alpha} = \frac{1}{2}  \mu_{\text{spin}}^{(m)} 
\langle Y_{2m} |  \hat{Q}_{\alpha \alpha} | Y_{2m} \rangle 
\end{math}
terms of the decomposition (\ref{tzspin}) for Co monolayers on Pd
(optimized geometry).  The sums of these components are shown in the
last row for each system and they correspond to the total \ms,
$T_{z}$, $T_{x}$, and $T_{y}$ of the $d$ electrons [evaluated using
  the approximative expression (\ref{tzspin}) in the case of 
  \Ta]. }
\label{tab-tzcomp}
\begin{ruledtabular}
\begin{tabular}{ldddd}
component & \mu_{\text{spin}}^{(m)} & 
  T^{(m)}_{z} &  T^{(m)}_{x} & T^{(m)}_{y} \\
\hline
  \multicolumn{5}{c}{Co on Pd(100)} \\ 
$xy$           & 0.319 &  0.092 & -0.046 & -0.046 \\
$yz$           & 0.465 & -0.066 &  0.133 & -0.066 \\
$3z^{2}-r^{2}$ & 0.365 & -0.104 &  0.052 &  0.052 \\
$xz$           & 0.465 & -0.066 & -0.066 &  0.133 \\
$x^{2}-y^{2}$  & 0.449 &  0.128 & -0.064 & -0.064 \\
sum            & 2.062 & -0.018 &  0.009 &  0.009 \\  [1.0ex]
  \multicolumn{5}{c}{Co on Pd(111)} \\ 
$xy$           &  0.339 &  0.097 & -0.048 & -0.048 \\
$yz$           &  0.428 & -0.061 &  0.122 & -0.061 \\
$3z^{2}-r^{2}$ &  0.490 & -0.140 &  0.070 &  0.070 \\
$xz$           &  0.428 & -0.061 & -0.061 &  0.122 \\
$x^{2}-y^{2}$  &  0.339 &  0.097 & -0.048 & -0.048 \\
sum            &  2.023 & -0.069 &  0.034 &  0.034 \\  [1.0ex]
  \multicolumn{5}{c}{Co on Pd(110)} \\ 
$xy$           &  0.397 &  0.113 & -0.057 & -0.057 \\
$yz$           &  0.346 & -0.049 &  0.099 & -0.049 \\
$3z^{2}-r^{2}$ &  0.515 & -0.147 &  0.074 &  0.074 \\
$xz$           &  0.527 & -0.075 & -0.075 &  0.151 \\
$x^{2}-y^{2}$  &  0.343 &  0.098 & -0.049 & -0.049 \\
sum            &  2.128 & -0.060 & -0.009 &  0.069
\end{tabular}
\end{ruledtabular}
\end{table*}

To get a more quantitative feeling of how the various contributions
add together to generate \Ta, we present in Tab.~\ref{tab-tzcomp} the
$m$-decomposed magnetic moment $\mu_{\text{spin}}^{(m)}$ and
individual terms of the sum (\ref{tzspin}) for Co monolayers on Pd
surfaces.  One can see that the $T_{\alpha}$ term is formed by a
competition between those $m$ components which contain the $\alpha$
coordinate and those which do not (they contribute with an opposite
sign, as it can be seen also from Tab.~\ref{tab-qaa}).  In fact, this
is what is meant by the statement that the \Ta\ term describes the
anisotropy of \ms.

Eq.\ (\ref{tzspin}) gives an intuitive insight into $T_{\alpha}$
provided that the underlying approximations --- the neglect of the
spin-orbit coupling and of the non-diagonal terms shown in 
Eq.~(\ref{nondiag}) --- are not too crude.  To check this, we compare
the values of \Ta\ calculated via the exact relation in
Eq.~(\ref{exact}) and via the approximative Eq.~(\ref{tzspin}).
Special attention is paid to the differences between the \Ta\ terms
for different orientations of \MM, because the
$7(T_{\alpha}-T_{\beta})$ quantities determine the apparent anisotropy
of \ms\ as deduced from the XMCD sum rule in Eq.~(\ref{eq-spin}).  The
outcome for both monolayers and adatoms is summarized in
Tab.~\ref{tab-tzterm}.  Let us recall that for bulk hcp Co, the
magnetic dipole term is very small (we get \tz=$-0.002$~\mB).  
Note that all
values presented in Tabs.~\ref{tab-tzcomp}--\ref{tab-tzterm} were
obtained from fully relativistic calculations, including the
spin-orbit coupling.

% Wide table planned for two columns
\begin{table*}
\caption{Magnetic dipole term for Co monolayers and adatoms on
  Pd(100), Pd(111) and Pd(110) (optimized geometries) for different
  magnetization directions.  For each system, the first line
  (``exact'') contains values calculated using Eq.~(\ref{exact}) and
  the second line (``approx.'') contains values calculated using
  Eq.~(\ref{tzspin}).  The $T_{y}$ terms were evaluated only for the
  (110) surface.  }
\label{tab-tzterm}
\begin{ruledtabular}
\begin{tabular}{llddddd}
   & & 
    \multicolumn{1}{c}{$T_{z}$}  &
    \multicolumn{1}{c}{$T_{x}$}  &
    \multicolumn{1}{c}{$T_{y}$}  &
    \multicolumn{1}{c}{$7(T_{x}-T_{z})$}  &  
    \multicolumn{1}{c}{$7(T_{y}-T_{z})$}  \\
\hline
  \multicolumn{7}{c}{Co on Pd(100)} \\ 
monolayer & exact &
  -0.017 & 0.010  &  & 
     0.188 &   \\
          & approx.\ &
  -0.018  & 0.009 &  &
     0.184 &    \\  [0.5ex]
adatom & exact &
  -0.024  & 0.015  &  &
     0.275 &  \\
       & approx.\ &
  -0.026  & 0.013  &  &
     0.276  &  \\  [1.0ex]
  \multicolumn{7}{c}{Co on Pd(111)} \\ 
monolayer & exact &
  -0.066  & 0.035  &  & 
     0.707  &  \\
          & approx.\ &
  -0.069 & 0.034  &  & 
     0.723  &  \\ [0.5ex]
adatom & exact &
  -0.146 & 0.080  &  &
     1.577 & \\
       & approx.\ &
  -0.154 & 0.077 &  &
     1.618  &  \\ [1.0ex]
  \multicolumn{7}{c}{Co on Pd(110)} \\ 
monolayer & exact &
  -0.057 &  -0.008 &  0.068  &
     0.339  & 0.872 \\ 
          & approx.\ &
  -0.060  & -0.009  &  0.069  &
     0.360  &  0.904  \\ [0.5ex]
adatom & exact &
  -0.112  & -0.020 &  0.141  &
     0.644  &  1.768  \\
         & approx.\ &
  -0.117  & 0.011 &  0.106  &
     0.900  &  1.566 \\
\end{tabular}
\end{ruledtabular}
\end{table*}

One can see from our results that the approximative expression for
\Ta\ works quite well for the Co-Pd systems: quantitative deviations
sometimes occur but the main trend is well maintained.  One can expect
that for systems with a strong spin-orbit coupling the deviations
between Eqs.~(\ref{exact}) and (\ref{tzspin}) will be larger.

The last two columns of Tab.~\ref{tab-tzterm} contain the values of
$7(T_{x}-T_{z})$ and, for the case of the (110) surface, also of
$7(T_{y}-T_{z})$.  These values are comparable to \ms\ which means
that even though \ms\ practically does not depend on the magnetization
direction at all, its combination \mm{\mu_{\text{spin}} +
  7T_{\alpha}}\ probed by the XMCD sum rule may strongly depend on the
magnetization direction.

%%%%%%%%%%%%%%%%%%%%%%%%%%%%%%%%%%%%%%%%%%%%%%%%%%%%%%%%%%%%%%%

\section{Discussion}   \label{sec-discuss}

We investigated how the magnetic properties of Co adatoms and
monolayers can be manipulated by selecting different supporting
Pd surfaces.  We found that this has a moderate effect on \ms\ and
\nh, larger effect on \mo\ and dramatic effect on the MAE and on the
\Ta\ term. For the adatoms the effect is larger than for the
monolayers.  Moreover, the transition from monolayers to adatoms has a
larger effect than a moderate variation in the height of the Co layer
above the substrate.  If the spin-orbit coupling is not very strong,
the \Ta\ term can be understood as arising from a competition between
those $m$-decomposed components of \ms\ which are associated with the
$\alpha$\ coordinate and those which are not.

In the past, the influence of the orientation of superlattices
(multilayers) on magnetic properties was already investigated,
however, the focus was mainly on the role of defects and interface
abruptness.\cite{LFL+90} Here, we deal with perfect monolayers and
surfaces and investigate how sole selection of a different surface can
affect various quantities related to magnetism.  Likewise, the
importance of the \tz\ term for an XMCD sum rules analysis has been
highlighted before when it was found that the absolute value of
\mm{7T_{z}}\ amounts to about 20~\% of \ms\ for some low-dimensional
systems\cite{KEDF02} or that for atomic clusters \ms\ can show a
different behavior with changing cluster size when compared to
\mm{\mu_{\text{spin}}+7T_{z}}.\cite{SME09b} In this study the
importance of the {\it anisotropy} of the magnetic dipole term in
nanostructures is stressed for the first time and it should be noted
that the anisotropy of \Ta\ which we highlight here is primarily
connected with the breaking of the crystal symmetry at the surface and
occurs even without spin-orbit coupling.

For the monolayers, the changes in \ms\ when going from one surface to
another reflect the corresponding changes in the coordination numbers:
\ms\ is largest for the (110) monolayer where each Co atom has got
only two nearest neighboring Co atoms, next comes the (100) monolayer
with four Co neighbors and the lowest \ms\ is obtained for the (111)
monolayer with six Co neighbors.  This complements an analogous trend
found earlier for free\cite{SKE+04} and supported
clusters.\cite{MLZ+06,SBM+07,BSM+12} The magnetic moments induced at
individual Pd atoms are larger for Co monolayers than for Co adatoms,
which reflects the fact that for monolayers, Pd atoms are polarized by
more than one Co atom.

% Figure planned for 1 column
\begin{figure}
\includegraphics[viewport=0.0cm 0.5cm 8.4cm 8.0cm]{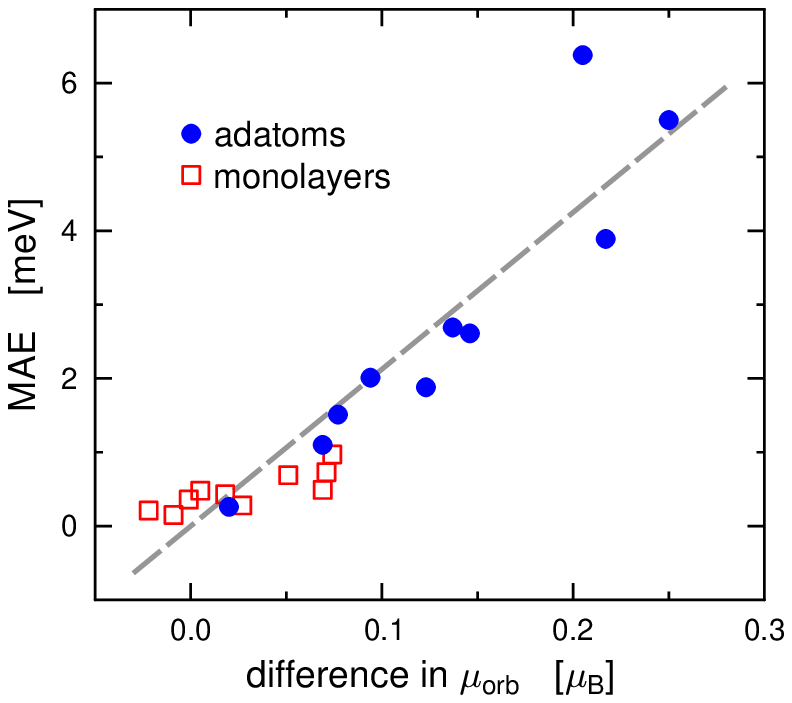}%
\caption{(Color online) Dependence of the MAE for Co monolayers and
  adatoms on the difference of orbital magnetic moments for respective
  magnetization directions.  The dashed line represents Bruno's
  formula in Eq.~(\protect\ref{bf}).}
\label{fig-bruno}
\end{figure}

The large amount of data gathered here for quite a complete set of
systems allows a comprehensive look at the relation between the MAE
and the anisotropy of \mo.  In this respect Bruno's
formula\cite{Bru89}
\begin{equation}
E^{(\alpha)} - E^{(\beta)} \: = \: - \frac{\xi}{4} \, 
\left[ \mu_{\text{orb}}^{(\alpha)} -  
 \mu_{\text{orb}}^{(\beta)} \right]
\label{bf}
\end{equation}
connecting the differences of total energies to the differences of
orbital magnetic moments for two orientations of the magnetization,
$\alpha$ and $\beta$, proved to be very useful\cite{DGvdL+97} despite
its limitations,\cite{RKF+01} which become more severe in the case of
multicomponent systems with large spin-orbit coupling parameter $\xi$
for the non-magnetic component.\cite{ASE07,SMME08} To assess the
situation for 3$d$-4$d$ alloys, we compare the differences $\Delta
\mu_{\text{orb}}$ and $\Delta E$, using all the appropriate values
given in Tab.\ \ref{tab-moms}.  The outcome is shown in
Fig.\ \ref{fig-bruno}, together with a straight line representing
Eq.\ (\ref{bf}).  Here we take 85~meV for the spin-orbit coupling
parameter $\xi$ [which appears to be a rather universal value for Co
  as our calculations yield $\xi$ of 85.4~meV, 84.5~meV, 84.9~meV and
  85.1~meV for bulk hcp Co and for a Co monolayer on Pd(100), Pd(111)
  and Pd(110), respectively].  It follows from Fig.\ \ref{fig-bruno}
that Bruno's formula Eq.~(\ref{bf}) works quite well for adatoms
(albeit with some ``noise'') but not so well for monolayers, where
relying solely on Eq.\ (\ref{bf}) might even lead to a wrong sign of
the MAE.  This may be connected with the fact that for monolayers, the
MAE is generally not very large and hence small absolute deviations
from the rule given in Eq.~(\ref{bf}) can lead to large relative
errors.

The sizable intraplanar anisotropy \mm{E^{(x)}-E^{(y)}}\ which we get
for a Co monolayer on Pd(110) had to be expected as this system could
be viewed as a set of Co wires which are surely anisotropic in this
respect.  However, we get a very strong azimuthal dependence of the
MAE also for the {\it adatom} on the (110) surface which is quite
surprising as this can be only caused by the underlying substrate.
The magnetic moments at Pd atoms are not very large
(Tab.\ \ref{tab-induced}), neither is the spin-orbit coupling
parameter $\xi$ for Pd in comparison to, say, 5$d$ elements. Thus,
this seems to be yet another example of the extreme sensitivity of the
MAE.  At the same time, let us note that the calculated azimuthal
dependence of the MAE can be accurately fitted by smooth sinusoidal
curves (see Fig.\ \ref{fig-azimuth}) which indicates a very good
numerical stability of the computational procedure.

The intraplanar anisotropy for a Co adatom on the Pd(111) surface can
be compared to similar systems investigated in the past.  In
particular, for a Co adatom on Pt(111) the amplitude of the
\mm{E^{(\|)}(\phi) - E^{(z)}}\ curve is about 2~\% of the average
value,\cite{BMS+07} i.e., similar to the current case.  For a
2$\times$2 surface supercell coverage of Fe on Pt(111), this amplitude
is 10--25~\% (depending on the geometry relaxation)\cite{THO+07} but
this situation is already quite distinct from the isolated adatom
case.

According to our calculations, a Co monolayer on Pd(100) has an
in-plane magnetic easy axis, a Co monolayer on Pd(111) has an
out-of-plane magnetic easy axis and the difference between the
respective MAE values is about 1~meV, which can be seen as a measure
of how much the out-of-plane magnetization is preferred by the
Co/Pd(111) system in comparison with the Co/Pd(100) system.  This is
similar to what was calculated for Co/Pd multilayers: both
Co$_{1}$Pd$_{3}$\ (100) and Co$_{1}$Pd$_{2}$\ (111) multilayers have
an out-of-plane magnetic easy axis but the MAE per unit cell is by
about 0.9~meV larger for the (111) multilayer than for the (100)
multilayer.\cite{KHY+96}

The theoretical values for the anisotropy of \Ta\ shown in
Tab.\ \ref{tab-tzterm} can be compared with experimental data for a
similar system, namely, a single Co(111) layer sandwiched between two
thick Au layers. By extrapolating results obtained via angle-dependent
XMCD measurements, Weller~\ea\cite{WSN+95} obtained
$7T_{x}=0.43$~\mB\ and $7T_{z}=-0.86$~\mB.  Our values for a Co
monolayer on Pd(111), $7T_{x}=0.24$~\mB\ and $7T_{z}=-0.46$~\mB\ (see
Tab.\ \ref{tab-tzterm}), are fully consistent with this.

We expect that our values for \mo\ will be systematically smaller than
experimental values because we rely in the LSDA which usually
underestimates \mo.\cite{HTW+96,CMK+08} The same may be also true for
the MAE. However, this does not affect our conclusions.

We used potentials subject to the ASA which may limit the numerical
accuracy of our results, particularly as concerns the MAE.  On the
other hand, our results do not differ too much from results of
full-potential calculations, especially in the case of monolayers.
For a Co monolayer on Pd(100), we get an in-plane magnetic easy axis
with an MAE of \mbox{-0.73}~meV per Co atom while Wu~\ea\cite{WCF+97}
obtained for the same $z_{\text{Co-Pd}}$ distance (1.65~\AA) a
theoretical MAE of \mbox{-0.75}~meV.  Magneto-optic Kerr
measurements\cite{MPE+07} as well as XMCD experiments\cite{STF+11}
showed that the magnetic easy axis of ultrathin Co films on
Pd(100) is indeed in-plane (the experiment includes also an in-plane
contribution from the shape anisotropy).  Note that the theoretical
MAE of \mbox{-0.18}~meV given in Ref.\ \onlinecite{MPE+07} was
obtained for a partially disordered Co monolayer simulating the growth
conditions, so it cannot be directly compared to our results obtained
for an ideal monolayer.

For a Co monolayer on Pd(111), we get a \ms\ value of 2.01~\mB\ in a
Co ASA sphere with a radius of 1.46~\AA\ while the full-potential
calculations of Wu \ea\cite{WLF+91} led to a \ms\ value of
1.88~\mB\ obtained within a Co muffin-tin sphere with a radius of
1.06~\AA\ --- both calculations thus again give consistent results.
For Pd atoms just below the Co layer, we get a \ms\ value of
0.32~\mB\ in a sphere with a radius of 1.49~\AA\ while the
corresponding \ms\ value of Wu \ea\cite{WLF+91} obtained within a
sphere having a radius of 1.32~\AA\ is 0.37~\mB.  In this last case,
one has to bear in mind that Wu \ea\cite{WLF+91} used a thin slab of
only five Pd layers sandwiched between two Co layers which clearly
favors a larger Pd polarization in comparison with just a single Co-Pd
interface considered in this work.

For adatoms, the ASA may be more severe than for monolayers,
nevertheless, the agreement between our calculations and the results
obtained via a full potential calculation is pretty good (see the end
of the Appendix).  As a whole, the accuracy of our calculations is
sufficient to warrant the conclusions which rely on comparing a large
set of data and not only on results for a singular system.

It follows from our results that one can change the magnetic easy axis
from in-plane to out-of-plane direction just by using as a substrate
another surface of the same element.  This could be used as yet
another ingredient for engineering the MAE of nanostructures, which
has become a great challenge recently.\cite{OVR+12} We also showed
that the magnetic dipole \Ta\ term can mimic a large anisotropy of
\ms\ as determined from the XMCD sum rules.  Hence, the anisotropy of
\Ta\ has to be taken fully into account when analyzing XMCD
experiments on nanostructures.

%%%%%%%%%%%%%%%%%%%%%%%%%%%%%%%%%%%%%%%%%%%%%%%%%%%%%%%%%%%%%%%

\section{Conclusions}   \label{sec-zaver}

Co monolayers and adatoms adsorbed on different surfaces of Pd
exhibit quite different magnetic properties.  The effect on \ms\ is
moderate, the effect on \mo\ is larger while the effect on the MAE and
on the magnetic dipole term \Ta\ may be crucial.  A surprisingly
strong azimuthal dependence of the MAE is predicted for a Co adatom on
Pd(110).

The dependence of \Ta\ on the direction of the magnetization can lead
to an apparent anisotropy of the spin magnetic moment as deduced from
the XMCD sum rules.  For systems with small spin-orbit coupling, the
\Ta\ term can be related to the differences between components of the
spin magnetic moment associated with different magnetic quantum
numbers.

%%%%%%%%%%%%%%%%%%%%%%%%%%%%%%%%%%%%%%%%%%%%%%%%%%%%%%%%%%%%%%%

\begin{acknowledgments}
This work was supported by the Grant Agency of the Czech Republic
within the project 108/11/0853, by the Bundesministerium f\"{u}r
Bildung und Forschung (BMBF) Verbundprojekt
R\"ontgenabsorptionsspektroskopie (05K10WMA) and by the Deutsche
Forschungsgemeinschaft (DFG) via SFB 689.  Stimulating discussions
with P.~Gambardella are gratefully acknowledged. 
\end{acknowledgments}

%--%--%--%--%--%--%--%--%--%--%--%--%--%--%--%--%--%--%--%--%--

\appendix* 

\section{Effect of the size of the relaxation zone}  \label{sec-size}

When studying the magnetism of adatoms, one should address the
question to which extent the host around the adatom has to be allowed
to polarize.  Zeller showed\cite{Zel93} that the polarization cloud
around a magnetic impurity in bulk Pd extends at least up to 1000
atoms.  \v{S}ipr \ea\cite{SBME10} showed that the convergence of the
MAE with respect to the slab thickness and/or with respect to the size
of the supercell which simulates the adatom is much slower than the
convergence of magnetic moments.  In view of these facts, it is
desirable to explore more deeply the situation for the systems
considered in this work.

As a test case, we select a Co adatom on Pd(111).  To facilitate the
comparison with calculations done by other methods, we put the Co
adatom in an hcp hollow site, with the vertical distance between the
Co adatom and the Pd surface layer as \zcp=1.64~\AA.  Our system is
thus similar to the system investigated by B{\l}o\'{n}ski
\ea\cite{BLD+10} (the main difference with respect to
Ref.\ \onlinecite{BLD+10} is that we do not consider any buckling of
the substrate).  To check the convergence with respect to the size of
the zone where the electronic structure is relaxed, we probed a series
of embedded cluster sizes, starting with relaxing the electronic
structure just in three Pd atoms (i.e., up to the distance of
2.3~\AA\ from the Co adatom) and ending with relaxing it in 220 Pd
atoms (up to 11.7~\AA\ from the Co adatom).  To safely accommodate
this large embedded clusters, we model the Pd substrate by a slab of
19 layers [contrary to 13 layers used in other calculations involving
  the Pd(111) surface in this work].  The largest embedded cluster
    with 220 Pd atoms contains Pd atoms located within the fith layer
    below the surface and comprises 329 sites altogether.

% Figure planned for 1 column
\begin{figure}
\includegraphics[viewport=0.3cm 0.3cm 8.5cm 7.7cm]{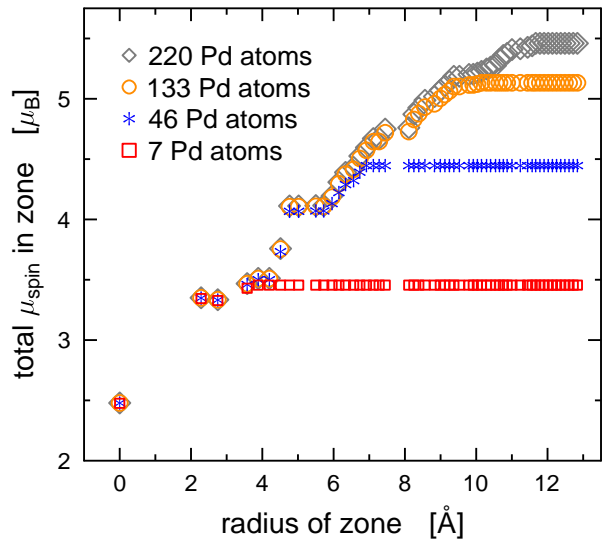}%
\caption{(Color online) Sum of the spin magnetic moments at the Co
  adatom and at those substrate Pd atoms which are enclosed in
  hemispherical zones of the given radii, for four embedded cluster
  sizes (identified by numbers of Pd atoms contained in them).}
\label{fig-spinconv}
\end{figure}

First we investigate the convergence of the spin magnetic moments.
This can be achieved by inspecting the total \ms\ contained inside a
hemisphere stretching from the adatom up to a certain radius.  The
dependence of this total \ms\ on the radius of the hemisphere forms an
``integral magnetic profile''.  This is presented in
Fig.\ \ref{fig-spinconv} for four embedded cluster sizes containing 7,
46, 133, and 220 Pd atoms, respectively.  The total \ms\ for a sphere
with zero radius is obviously just the \ms\ value of the Co adatom.
With increasing sphere radius the spin magnetic moments of enclosed Pd
atoms are added to it.  If the radius of the hemisphere becomes larger
than the radius of the embedded cluster, the total \ms\ obviously does
not change any more because the Pd atoms outside the embedded impurity
cluster are nonmagnetic.

It follows from Fig.\ \ref{fig-spinconv} that the spin magnetic moment
of the adatom as well as magnetic moments induced in the nearest Pd
atoms are actually already well described by relatively small embedded
clusters. However, the total \ms\ converges only very slowly with
increasing size of the relaxation zone because even quite distant Pd
atoms still contribute with their non-zero \ms.  Our results suggest
that the magnetic moments on all the Pd atoms do not arise due to a
direct interaction with the Co adatom.  Rather, the adatom induces a
magnetization in its nearest neighbors, then these further induce
magnetization in the next coordination shell and so on.  The emerging
picture of how the magnetism spreads through the Pd host is thus
consistent with the picture suggested by Polesya \ea\cite{PMS+10} in
terms of an exchange-enhanced magnetic susceptibility (see Fig.~4 of
Ref.\ \onlinecite{PMS+10} and the associated text).  A plot analogous
to Fig.\ \ref{fig-spinconv} could also be drawn for \mo\ exhibiting
the same features as seen in Fig.\ \ref{fig-spinconv}.
 
% Figure planned for 1 column
\begin{figure}
\includegraphics[viewport=0.0cm 0.5cm 8.4cm 8.0cm]{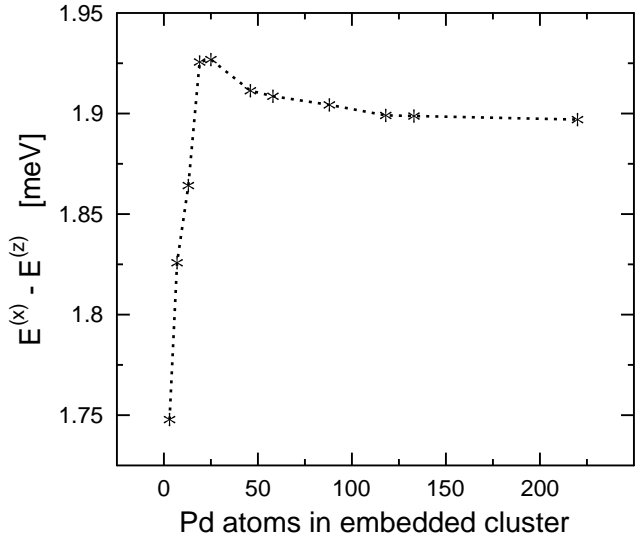}%
\caption{The MAE of a Co adatom in an hcp position on Pd(111) for
  different sizes of the embedded clusters.}
\label{fig-maeconv}
\end{figure}

Our results on the convergence of the magnetic moments may raise
objections about the convergence of the MAE.  If embedded clusters
containing as much as 220 Pd atoms still do not fully account for the
host polarization, can one get reliable results for the MAE, which is
sensitive to the way the substrate is treated?\cite{SBME10} To check
this, we calculated the MAE for a series of embedded cluster sizes
(Fig.\ \ref{fig-maeconv}).  One can see that in fact the MAE converges
quickly with increasing size of the embedded cluster. Already with a
relaxation zone including only 46 Pd atoms, which corresponds to a
radius of the hemisphere of 6.9~\AA\ containing Pd from up to the
third Pd layer below the surface, the accuracy of the MAE is better
than 1~\%.  This means that all the results presented in this work are
well converged.

The data in Fig. \ref{fig-maeconv} demonstrate that it is sufficient
to include a rather small polarization cloud within the Pd host in
order to get convergence in the MAE values.  More distant Pd atoms do
not contribute to the MAE, {\it even if they are magnetically
  polarized}.  This conclusion is not in contradiction with an earlier
result that reliable values of the MAE can be obtained only if the
host is represented by slabs of at least ten layers\cite{SBME10}
because that result concerned the total ``physical'' size of the model
system while in this appendix we focus only on the size of the zone
where the electronic structure is allowed to relax to the presence of
an adatom (or of an adsorbed monolayer).

To complete this part, we should compare our results with the results
of B{\l}o\'{n}ski \ea\cite{BLD+10} which were obtained by performing a
plane-wave projector-augmented wave (PAW) calculation for a supercell
comprising five-layers thick slabs and a 5$\times$5 surface
unit cell.  As concerns the Co adatom itself, \ms\ and \mo\ for the
in-plane magnetization direction and \mo\ for the out-of-plane
magnetization direction are 2.48~\mB, 0.15~\mB, and 0.27~\mB\ in this
work and 2.24~\mB, 0.19~\mB, and 0.22~\mB\ in B{\l}o\'{n}ski
\ea\cite{BLD+10} As concerns the MAE calculated via the magnetic force
theorem (torque method), it is 1.90~meV out-of-plane in this work and
0.72~meV out-of-plane in B{\l}o\'{n}ski \ea\cite{BLD+10} The value for
the induced \ms\ in the nearest Pd atoms is 0.28~\mB\ in this work and
0.33~\mB\ in B{\l}o\'{n}ski \ea\cite{BLD+10} All these values
are in good agreement, considering the differences between both
approaches.

% Produces the bibliography via BibTeX.

%%%\bibliography{liter_co-on-pd}

% File *.bbl inserted manually in order to avoid need for BibTeX
% cooperation;  this is an aid to PR publishing


\end{document}